# SENTIMENT ANALYSIS OF DOCUMENT BASED ON ANNOTATION

Archana Shukla

Department of Computer Science and Engineering, Motilal Nehru National Institute of Technology, Allahabad
archana@mnnit.ac.in

## ABSTRACT

*I present a tool which tells the quality of document or its usefulness based on annotations. Annotation may include comments, notes, observation, highlights, underline, explanation, question or help etc. comments are used for evaluative purpose while others are used for summarization or for expansion also. Further these comments may be on another annotation. Such annotations are referred as meta-annotation. All annotation may not get equal weightage. My tool considered highlights, underline as well as comments to infer the collective sentiment of annotators. Collective sentiments of annotators are classified as positive, negative, objectivity. My tool computes collective sentiment of annotations in two manners. It counts all the annotation present on the documents as well as it also computes sentiment scores of all annotation which includes comments to obtain the collective sentiments about the document or to judge the quality of document. I demonstrate the use of tool on research paper.*

## KEYWORDS

*Sentiment Analysis, Opinion Mining, Classification, Annotation, Sentiment Words, Polarity Value*

## 1. INTRODUCTION

Several degree programs of universities academic institute have research component of varying duration from six months to four-five years. As a first activity in the research, students are advised to survey literature related to their domain of interest to define their proposed activity. They collect research papers and other publication either from web sites of professional societies like IEEE, ACM, and LNCS or from printed copy of journals available in their library. While going through these research publications, they marks underline on some important parts, or they highlights some words or phrases or whole sentences or paragraph. They also write their notes, observations, remarks, questions etc either on the same document or on the separate sheet of paper. These highlights or underline or comments/observation may be about entire paper or part of them. At some point of time, they collate and integrate these observations to identify and define their research problems. On completions of their degree programs these knowledge represented in the form of observation and thoughts are lost as they are not saved and shared by the next batch of students.

These observation/comments/highlights/underline are very valuable knowledge resource not only for the current reader but also for future generation of students who are likely to work in the same area. However, at present these knowledge resources are not available to future generation as they are not available in electronic form and are not sharable.

My work is motivated by desire to provide a tool which provides a facility to record their comments, notes, observation, and explanation , highlights, underline etc. either on document or on another comments and evaluate the collective sentiments of the researchers over the document. These collective sentiments of annotators may be used as an indicator of quality or usefulness of the documents.

I have developed a tool, KMAD [16], which provides facility to annotate either PDF documents or another annotation. It also creates knowledgebase consisting of annotations and metadata found in the document.

In this paper, I describe augmentations made in KMAD for the analysis of annotations found in the document to infer collective sentiments of annotators to judge the quality of document or its usefulness. The relationships between annotations are complex. Meta annotation, which is annotation over another annotation, may be comment on annotation which is of type comment, note, explanation, help etc. We only consider those annotation or Meta annotations which are of type comment, highlights and underline.

Collective sentiments of annotators are visualized either in terms of positive sense, or negative sense or neutral sense based on adjectives, adverbs, verbs etc using *WordNet*. These collective sentiments of annotators may be used as a characteristic features to judge the quality or usefulness of the document.

My tool uses *SentiWordNet* to assigns sentiment scores to each word found in annotations. Sentiments of words are assigned three sentiment scores: positivity, Negativity and objectivity with a word and lies in between the range of [0-1].

This paper is organized in seven sections. Section 2 presents the related work. Section 3 briefly describes the services provided by KMAD without augmentation. In Section 4, we describe opinion analyzer. In section 5, we describe the augmented KMAD tool. In section 6, we present the design and implementation details of our application and last section 7 present the conclusion.

## 2. RELATED WORK

Several research efforts have been made to analyze documents, highlights, underline, comments, annotations on document and web sites to evaluate collective sentiments of readers or evaluators. [1] [2] considered contents of the documents for analysis of sentiments. They extracted the sentiment words consisting of adjectives and nouns using GI (*General Inquirer*) and *WordNet*. They assigned sentiment value to each extracted word based on number of times it appears in the whole document. They assigned sentiment polarity using sentiment lexicon database which include 2500 adjectives and 500 nouns where for each word, sentiment definition was defined in terms of (word, pos, sentiment category). Jiang Kim & Hovy and Weibe & Riloff [3, 4] analyzed the text file related to a given topic. They they used their own dictionary which included 5880 positive adjectives, 6233 negative adjectives, 2840 positive verbs and 3239 negative verbs instead of considering generalized ontology for extracting sentiment words. For unseen word, they assigned sentiment strength by computing probability of word based on count occurrences of word synonyms in the dictionary. YANG et al [5] did analysis of online document of Chinese review based on topic. Topic of a review was identified using n-gram approach. Sentiment words were extracted using four dictionaries such as Positive Word Dictionary, Negative Word Dictionary, A student Positive and Negative Word Dictionary and HowNet. Polarity values were assigned by computing average score based on term frequency of word. Positive and Negative value of words were assigned manually by annotators.

[6, 7, 8, 9, 10] extracted the sentiment words consisting of adjectives or adverbs or adjective-adverb both. They proved that subjectivity of a sentence could be judged according to the adjectives/adverb in it. Polarity value for each word was assigned by calculating the probability based on term frequency of word.

Several authors also did classification of document based on annotation. Emmanuel Nauer et al [11] did classification of the HTML document based on the annotation. The content of the document is annotated and similarity will be matched based on the domain Ontology. Michael G. Noll et al [12] did classification of the web document by analyzing the large set of real world data. They interested to find out what kinds of documents are annotated more by the end-users. Anotrea Mazzei [13] did classification of extracted handwritten annotations on machine printed documents based on type of annotations such as underline annotation, highlighted annotation, annotations in margins and blank space, and annotation in between the lines or over the text. Steimle et al [15] developed a system *CoScribe* which provided facility to annotate and classify power point lectures slides based on four types of annotation such as important, to do, question and correction. Sandra Bringay et al [16] did classification of the electronic health record based on both informal and formal annotations for managing knowledge. They had designed a schema for formal annotation which includes author name, date time, place, document, target, annotation type. Annotation type consisted of comment, link between two documents, a message for a precise recipient, an annotation created in order to write a synthesis, a response to a annotation. Informal annotation was used by practitioners when they would like to give some brief history about the patients or disease.

All the above works have contributed significantly in the field of sentiment analysis and classification of document in different domains. We have focused on research academy domain to analyze annotations on research papers to obtain the collective sentiments. Our sentiment words include adjectives, verbs, adverbs, nouns found in comments, highlights and underline. We expect that it will give better result because we give bigger set of sentiment words.

Our annotation schema has similarity and difference with the one used in [16]. Our comments reflect evaluative judgment of annotation whereas that of [16] summarizes the content. Our meta-annotation is very similar to [16].

## 3. KMAD

I have developed a tool named KMAD [14] to annotate a PDF document. We have designed an annotation schema using DTD (Document Type Definition) to capture the information of annotation.

```
<? xml version = "1.0" standalone ="yes">
<! DOCTYPE Annotations [
<! ELEMENT Annotation_List (Annotation)*>
<! ELEMENT Annotation (Author, Type, Annotation_on, Comment, Date_Time, Paper)>
<! ELEMENT Author (#PCDATA)>
<! ELEMENT Type (note | comment | help | insert)>
<! ELEMENT Annotation_on EMPTY>
<! ELEMENT Comment (type, comment) +>
<! ELEMENT type (note | comment| help| insert)>
<! ELEMENT comment (#PCDATA)>
<! ELEMENT Date_Time (#PCDATA)>
<! ELEMENT Paper (#PCDATA)>
<! ATTLIST Comment comment_id ID #REQUIRED >
<! ATTLIST Paper paper_id ID #REQUIRED>
<! ATTLIST Creator annotator_id ID #REQUIRED>
<! ATTLIST Annotation_on p_id IDREF>
<! ATTLIST Annotation_on c_id IDREF> ]>
```

These annotations contain *Author*, *Type*, *Annotation_on*, *Comment*, *Date_Time* as elements. *Annotation_id*, *PDF_Paper_id*, *Comment_id* as an attribute. The element *Type* may take either "*note*" or "*comment*" or "*help*" or "*insert*" or "*paragraph*" or "*unknown*" as values. *Note* type indicates that annotation summarizes the content whereas *comment* type indicates weather it is an evaluation or criticism of the content. My tool also captures the relationship between PDF documents and associated annotations and also between annotations and about annotations.

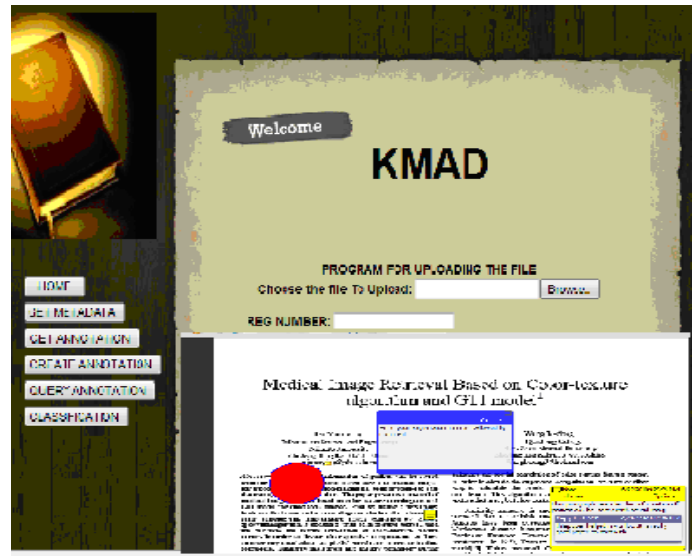

**Fig1.** Annotation Creation Process

It provides user friendly interface to upload a PDF document and to create annotation on document. Author can either visualize annotation along with document or only annotations. A snap shot for our tool is shown in Fig 1.

My application creates a relational database for annotations, PDF documents and relationship between them. It also keeps the record of each annotator. This database can be queried to find all annotations with a PDF document or to model the annotators. It also support search on PDF document based on any of the metadata or annotator_id.

## 4. OPINION ANLAYZER

My tool analyzes annotations to obtain the collective sentiments of annotators in two manners to judge the quality or usefulness of documents. In first method, it extracts all the annotations created by annotators which are of type highlights, underline, and comments and count total number of annotations present on the document. Its helps to infer that more number of annotations indicates good quality of document or its usefulness. My tool also gives facility to view all the annotations such as highlights, underline and comments also.

**Algorithm 1** Find the total count of annotation found in the PDF document.

**Input:** Annotated PDF document

**Output:** Total Count of Annotation present on the PDF Document
1. **for** each annotated PDF document **do**
2. get all annotation which is of type highlights, underline, comments present on the document
3. Count all annotation i.e. count= sum (Annotation1 + annotation2+ annotation3 +….+ annotation N)
4. **end for.**
5. Output Count.

In second method, it computes collective sentiment of annotators. It is a two step process. In first steps, it computes the average scores for all annotations. A detail of this process is given in Algorithm 2. In second steps, it computes the weighted average of score of annotation to infer the collective sentiments of author. A detail of this process is given in Algorithm 3. We have used SentiWordNet to assign polarity scores. Three scores are associated with each sentiment word in terms of positivity, negativity and objectivity.

**Algorithm 1** Find the average score of each annotation found in PDF document

**Input:** List of sentiment words extracted from comments of annotation

**Output:** Sentiment score
1. **for** each sentiment word from List **do**
2. Get polarity as well as sentiment scores using *SentiWordNet*.
3. **if** word is preceded with negation word "Not" **then,**
4. Interchange positive and negative sentiment scores of the word which comes after the Not.
5. Record above sentiment scores of each word in Table
4. end if
5. **end for**
6. Get maximum polarity value of each word from the table to compute average score.
7. **if** maximum polarity value is negativity **then** ,
8. Make the score negative.
9. Compute sentiment score such as S.S (sentiment Score) = add (maximum   polarity value of each word)/ Total number of words found in comment.
10. Output S.S (Sentiment score).

Algorithm 2 is used to find weighted average of sentiment score to infer the collective sentiments of annotator over the document. This algorithm takes input as sentiment score of each annotation and we also eliminate scores of annotation on which another annotation has contradicting sentiments.

**Algorithm 2** Find weighted average of sentiment score on PDF document

**Input:** sentiment score, number of meta-annotation on annotation

**Output:** weighted average score of each comment of annotation

1. **for** each sentiment score **do**
2. **if** sentiment score is of annotation on annotation **then**,
3. **if** sentiment score negates the previous sentiment score **then**,
4. Exclude that sentiment score from the computation.
5. **else**
6. Compute weighted average of sentiment score = sentiment score * number of meta-annotation/total number of annotation on a document.
7. **end if ;end if**
8. **end for**
9. **if** weighted average of sentiment score is positive **then**,
10. Result "sentiment of collective annotator over the document is positive".
11. **else**
12. Result "sentiment of collective annotator over the document is negative".
13. **end if**

We demonstrate our algorithm using example annotations on a document. These annotations are given as shown in Fig 2.

Ann1: This article is quite well but not so good.

Ann2: I am satisfy with this comment.

Ann3: It is not a good one.

Ann4: This is the best article.

Ann5: It is good article but not best one.

Ann6: It is bad one.

Ann7: Not best one but quite well.

**Fig. 2.** Annotations

List of sentiment words found in annotations of document as shown in Table 1.

**Table 1.** Words with associated sentiment scores and their maximum polarity

|       | Word   | Positive value | Negative value | Objective value | Max(polarity value) |
|-------|--------|----------------|----------------|-----------------|---------------------|
| Ann1  | Quite  | 0              | 0.625          | 0.375           | -0.625              |
|       | Well   | 0.75           | 0              | 0.25            | +0.75               |
|       | Not    | 0              | 0.625          | 0.375           | -0.625              |
|       | Good   | 0.875          | 0.125          | 0               | -0.875              |
| Ann2  | Satisfy| 0.5            | 0              | 0.5             | +0.5                |
| Ann3  | Not    | 0              | 0.625          | 0.375           | -0.625              |
|       | Good   | 0.875          | 0.125          | 0               | -0.875              |
| Ann4  | Best   | 0.75           | 0              | 0.25            | +0.75               |
| Ann5  | Good   | 0.875          | 0.125          | 0               | +0.875              |
|       | Not    | 0              | 0.625          | 0.375           | -0.625              |
|       | Best   | 0.75           | 0              | 0.25            | -0.75               |
| Ann6  | Bad    | 0              | 0.625          | 0.375           | -0.625              |
| Ann7  | Not    | 0              | 0.625          | 0.375           | -0.625              |
|       | Best   | 0.75           | 0              | 0.25            | -0.75               |
|       | Quite  | 0              | 0.625          | 0.375           | -0.625              |
|       | Well   | 0.75           | 0              | 0.25            | +0.75               |

Sentiment score of annotation 1= ((-0.625)+(0.75)+(-0.625)+(-0.875) /4) = -0.34375

Similarly, Average score of annotation 2 = +0.75

Total weighted average of sentiment score of document =  +0.29375

So, if the above value is positive then, sentiment of document is positive, otherwise negative. Here, sentiment of collective annotator over document is positive.

## 5. AUGMENTED KMAD TOOL

I have described augmentation made in KMAD tool [14] for the analysis of annotations as shown in Fig 3.

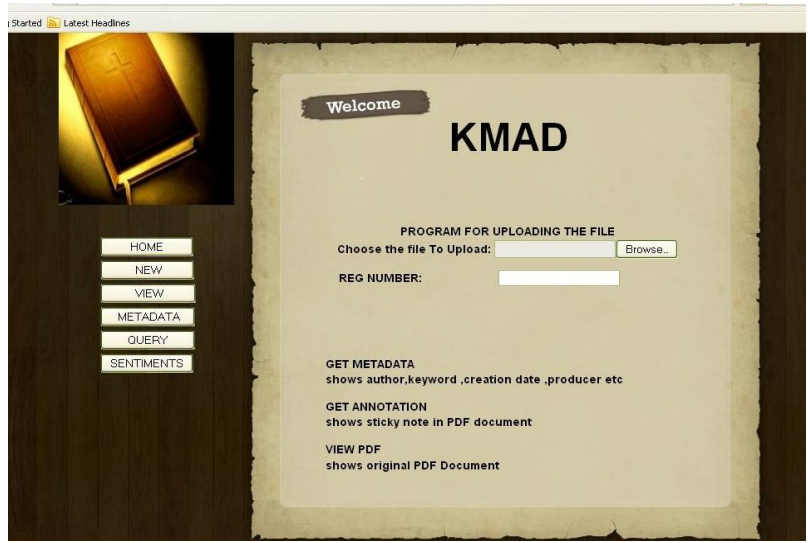

Fig 3. Home page

It provides five clickable buttons to perform different tasks. These tasks include *new*, *view* annotations such as comments, highlights, underline either page wise or collectively, *Metadata*, *Query, Sentiments.*

My application computes collective sentiment of annotators in two ways. In First method, it extracts all the annotations found in the PDF document such as comments, highlights, underline and count total number of annotations which helps to infer the usefulness of the document as shown in Fig 4.

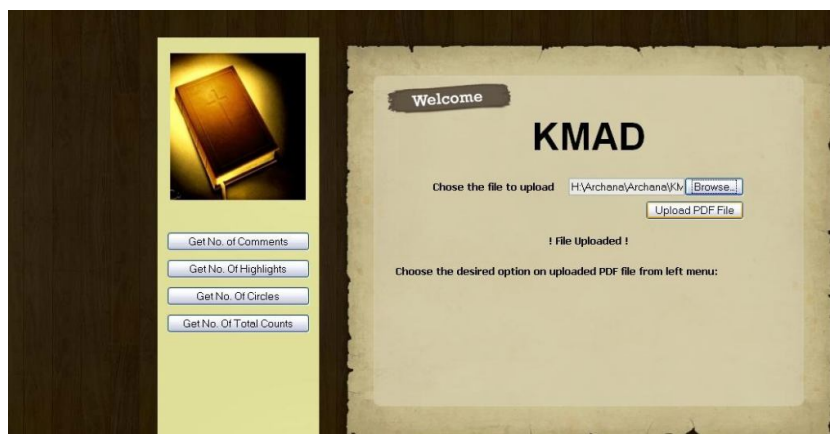

Fig 4. Sub-option available in view option

My application provide four functions under view option to view number of comments, number of highlights, number of underline and also the total counts of all annotations present in the document as shown in Fig 4. This result help researcher to infer the usefulness of document whether respective document is relevant to their area or not. This has been implemented using API of java PDFBOX.

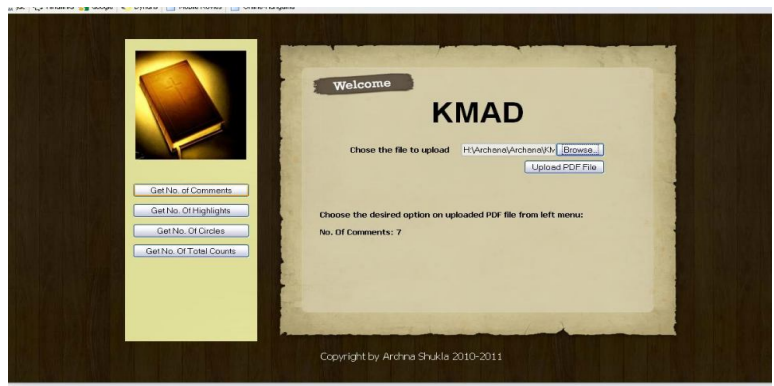

Fig 5. Total number of comments present in the document.

It allows user to upload annotated document to view either of these four sub option. Snapshot is given as in Fig5, Fog6, Fig 7, and Fig 8 respectively.

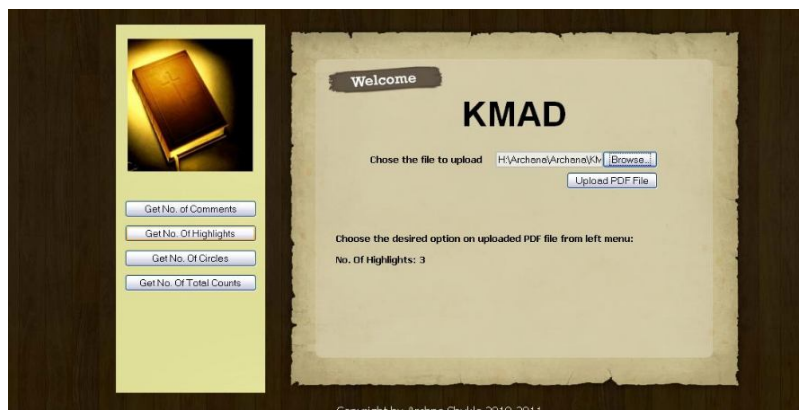

Fig 6. Total number of highlights present in the document.

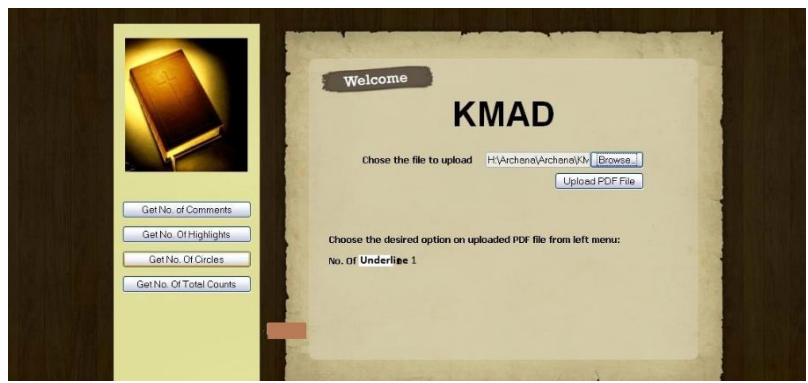

Fig 7. Total number of underline present in the document.

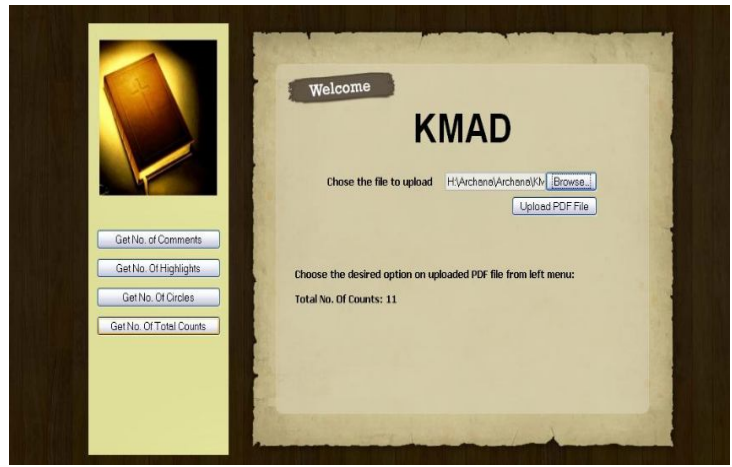

Fig 8. Total number of counts present in the document.

In second method, my application extracts the sentiment words such as adjectives, adverbs or verbs from annotation such as comment found in the document to evaluate the collective sentiments of annotators as shown in Fig 9. It also assigns sentiment scores to each sentiment words found in comments of annotations using *SentiWordNet* as shown in Fig 10.

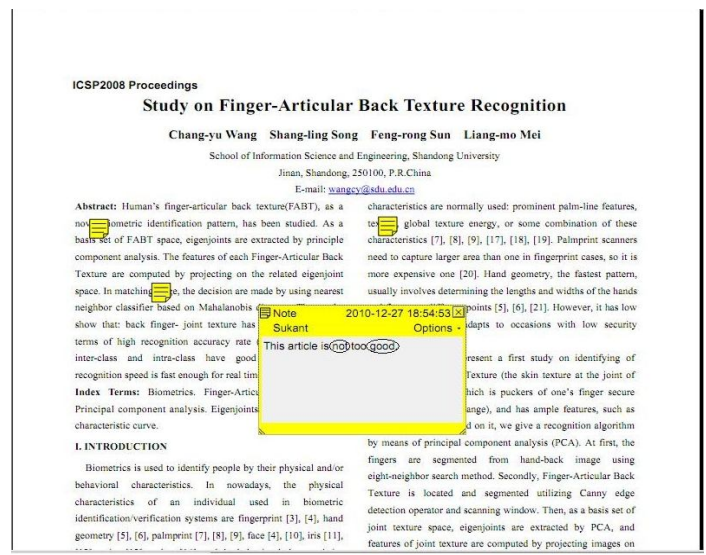

Fig 9 Annotator Comment

| Word → Polarity ↓ | Positive Polarity | Neutral Polarity | Negative Polarity |
|---|---|---|---|
| Not | 0.0 | 0.375 | 0.625 |
| Good | 0.875 | 0.125 | 0.0 |

Fig 10 Sentiment Scores for Sentiment Words

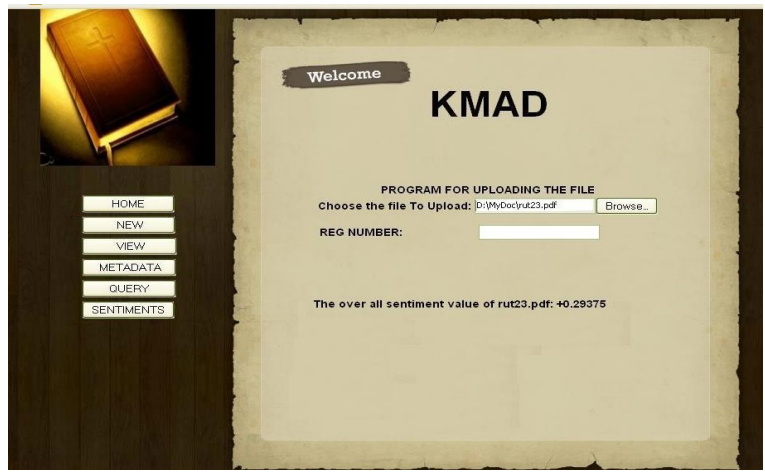

**Fig. 11** Collective sentiments over document.

My application computes total weighted average of sentiment score of annotations found in PDF document to infer the collective sentiments of annotators as shown in Fig 11. It also allows authors to view list of annotation available on PDF document created by them either page wise or collectively. Our application also extracts the metadata such as Title, Author, keywords, summary, date-time using function of PDF BOX API.

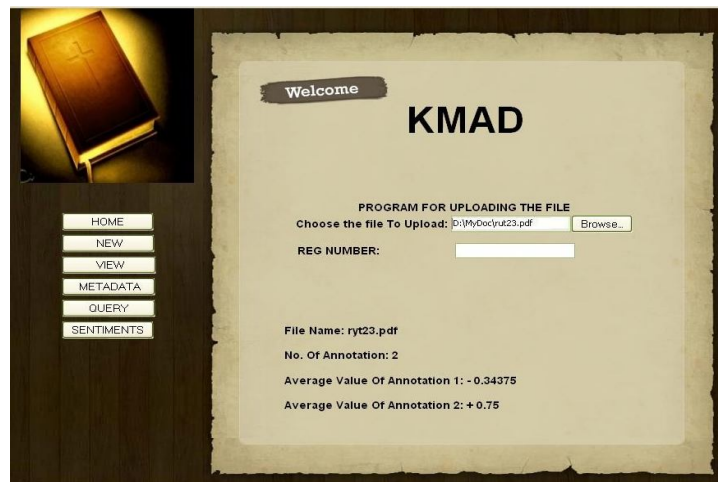

**Fig.12.** Sentiment scores of annotations

My application allows authors to query about sentiment score or collective sentiments of annotator which is available on PDF document either on the basis of annotator-id or on the basis of file name as shown in Fig 12.

## 6. DESIGN AND IMPLEMENTATION

I have used three layer architecture for our augmented KMAD tool. The top most layer is the presentation layer, which manages all the interaction to end user. The middle layer is the application logic layer which includes all the functionalities such as *annotation extractor module*, *sentiment word extractor module*, *SentiWordNet* and *WordNet* which are used to manage knowledge resources. The bottom layer is the database layer and contains the database for document, document Metadata, Annotation, annotation relation and sentiment words.

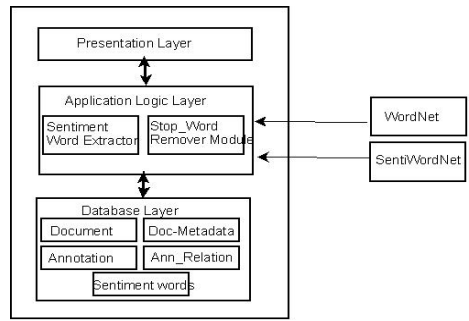

Fig 13 Augmented KMAD Architecture

My tool extracts the annotation using PDF BOX API such as *getDocumentCatalog()* for extracting the page information on which annotation has been done. Total number of pages in the PDF document and their count is listed using *getAllPages()* and *size()* function respectively. Extracts annotation field list available on a PDF document using *getAnnotation()* function. This function maintains the list of all annotation. If annotation field is of type "text", then , for each annotation field in the annotation list extracts information of annotation field such as comment using *getContents()*function. It also removes stop words and performs stemming using one of the module and consider adjectives, nouns, adverbs and verbs based on POS tagging using *WordNet*. It assigns polarity values in terms of positive, negative and objective using *SentiWordNet*.

Sentiment scores lie in between the range of [0.0-1.0]. At the time of assignment of scores, our tool also takes care of negation words such as "Not", "Never". If these words are found before any other word (*Adj*), then it interchanges +ve and –ve polarity values of that word which comes after "Not".

I have created a relational database as per ER diagram as shown in Fig 11. It shows the entity relationship between Document, Document metadata, Annotation, Annotation relation, Words and their associated sentiments. Our database contains five tables PDF_document, PDF_Annotation, Annotation_Annotation, Sentiments_Words, Sentiments_Annotation.

All the information extracted related to annotations and the relationship between annotations is available in a separate xml file. Our tool also stores all these information in relational database.

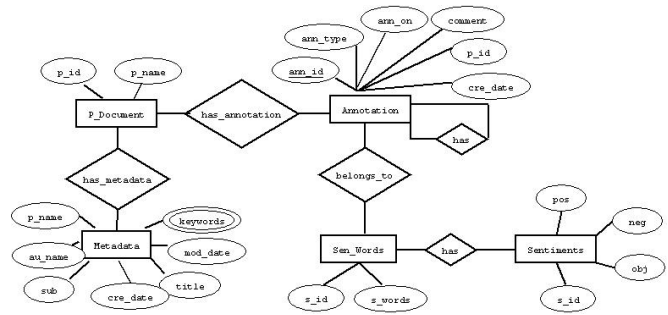

Fig 14 Entity Relationship Diagram

## 7 Conclusion

We have developed an augmented KMAD tool implemented using java server programming language to infer the collective sentiment of annotators and query knowledge base containing metadata , annotations and sentiments. We believe that it is helpful to research community. The relationship between the annotations is complex. We have only considered those annotations or meta annotation which is of type comment, highlights, and underline. In future we plan to consider more complex type relations.

## ACKNOWLEDGEMENTS

I would like to express my gratitude to Prof. B.D. Chaudhary, Motilal Nehru National Institute of Technology, Allahabad for the support and time he spent and discuss various aspects related to annotation.